\documentclass[]{aa}  
\usepackage{graphicx}
\usepackage{txfonts}
\begin{document} 

   \title{Age and helium content of the open cluster NGC 6791 from multiple eclipsing binary members.\thanks{Tables A1 and A2 are only available in electronic form at the CDS via anonymous ftp to cdsarc.u-strasbg.fr (130.79.128.5) or via http://cdsweb.u-strasbg.fr/cgi-bin/qcat?J/A+A/}}

   \subtitle{III. Constraints from a subgiant}

   \author{K. Brogaard
          \inst{1,2}\thanks{E-mail: kfb@phys.au.dk}
          \and
          F. Grundahl\inst{1,2}
          \and
          E. L. Sandquist\inst{3}
          \and
          D. Slumstrup\inst{4,1}
          \and
          M. L. Jensen\inst{5}
          \and
          J. B. Thomsen\inst{5}
          \and
          J. H. Jørgensen\inst{5}
          \and
          J.\,R.\,Larsen\inst{5}
          \and
          S. T. Bjørn\inst{5}
          \and
          C. T. G. Sørensen\inst{5}
          \and
          H. Bruntt\inst{1}
          \and
          T. Arentoft\inst{1}
          \and
          S. Frandsen\inst{1}
          \and
          J. Jessen-Hansen\inst{1}
          \and
          J. A. Orosz\inst{3}
          \and
          R.\,Mathieu\inst{6}
          \and
          A. Geller\inst{7,8}
          \and
          N. Ryde\inst{9}
          \and
          D. Stello\inst{10,11,1}
          \and
          S. Meibom\inst{12}
          \and
          I. Platais\inst{13}
          }

% List of institutions
\institute{
Stellar Astrophysics Centre, Department of Physics and Astronomy, Aarhus University, Ny Munkegade 120, DK-8000 Aarhus C, Denmark
\and
Astronomical Observatory, Institute of Theoretical Physics and Astronomy, Vilnius University, Saul\.{e}tekio av. 3, 10257 Vilnius, Lithuania
\and
Department of Astronomy, San Diego State University, San Diego, CA 92182, USA
\and
European Southern Observatory, Alonso de Córdova 3107, Vitacura, Santiago, Chile
\and
Department of Physics and Astronomy, Aarhus University, Ny Munkegade 120, 8000 Aarhus C, Denmark
\and
Department of Astronomy, University of Wisconsin-Madison, Madison, WI 53706, USA
\and
Center for Interdisciplinary Exploration and Research in Astrophysics (CIERA) and Department of Physics and Astronomy, Northwestern University, 1800 Sherman 
Avenue, Evanston, IL 60201, USA
\and
Adler Planetarium, Department of Astronomy, 1300 South Lake Shore Drive, Chicago, IL 60605, USA
\and
Department of Astronomy and Theoretical Physics, Lund Observatory, Lund University, Box 43, 221 00, Lund, Sweden
\and
School of Physics, University of New South Wales, NSW 2052, Australia
\and
Sydney Institute for Astronomy (SIfA), School of Physics, University of Sydney, NSW 2006, Australia
\and
Harvard-Smithsonian Center for Astrophysics, Cambridge, MA 02138
\and
Department of Physics and Astronomy, Johns Hopkins University, 3400 North Charles Street, Baltimore, MD 21218, USA
}

   \date{Received xx, 2021; accepted xx, 2021}

% \abstract{}{}{}{}{} 
% 5 {} token are mandatory
 
  \abstract
   {Models of stellar structure and evolution can be constrained using accurate measurements of the parameters of eclipsing binary members of open clusters. Multiple binary stars provide the means to tighten the constraints and, in turn, to improve the precision and accuracy of the age estimate of the host cluster.
In the previous two papers of this series, we have demonstrated the use of measurements of multiple eclipsing binaries in the old open cluster
NGC\,6791 to set tighter constraints on the properties of stellar models than was previously possible, thereby improving both the accuracy and precision of the cluster age. }
   {We identify and measure the properties of a non-eclipsing cluster member, V56, in NGC\,6791 and demonstrate how this provides additional model constraints that support and strengthen our previous findings.
}
   {We analyse multi-epoch spectra of V56 from FLAMES in conjunction with the existing photometry and measurements of eclipsing binaries in NGC6971.}
   {The parameters of the V56 components are found to be 
$M_{\rm p}=1.103\pm 0.008 M_{\odot}$ and $M_{\rm s}=0.974\pm 0.007 M_{\odot}$,
$R_{\rm p}=1.764\pm0.099 R_{\odot}$ and $R_{\rm s}=1.045\pm0.057 R_{\odot}$,
$T_{\rm eff,p}=5447\pm125$\,K and $T_{\rm eff,s}=5552\pm125$\,K,
and surface [Fe/H]=$+0.29\pm0.06$ assuming that they have the same abundance.}
   {The derived properties strengthen our previous best estimate of the cluster age of $8.3\pm0.3$ Gyr and the mass of stars on the lower red giant branch (RGB), which is $M_{\rm RGB} = 1.15\pm0.02\,M_{\odot}$ for NGC\,6791. These numbers therefore continue to serve as verification points for other methods of age and mass measures, such as asteroseismology. 
}

   \keywords{
   binaries:spectroscopic -- stars:fundamental parameters -- stars: individual: V604\,Lyr -- stars:abundances -- open clusters and associations: individual: NGC\,6791
            }

   \maketitle
%
%-------------------------------------------------------------------

\section{Introduction}

Star clusters allow age estimates of their stars from various ways of isochrone fitting to their photometric colour-magnitude diagram (CMD). If the cluster hosts eclipsing binary stars close to the turn-off, they can improve the age measurements because they provide direct measurements of mass and radius at age sensitive locations in the CMD. This has been exploited in both globular clusters (e.g. \citealt{Brogaard2017, Kaluzny2014,Kaluzny2015,Thompson2010, Thompson2020}) and open clusters (e.g. \citealt{Brogaard2011, Kaluzny2006, Knudstrup2020, Meibom2009, Sandquist2016, Torres2020}). However, sometimes the eclipsing systems found are on the main sequence, which still allows some model constraints, but not on the age \citep[e.g.][]{Brogaard2021}. 

The properties of the old open cluster NGC\,6791 were examined in detail by \citet{Brogaard2011,Brogaard2012}, who exploited the presence of two eclipsing systems, V18 and V20, at the cluster turn-off for a strong age constraint. Here, we demonstrate how a non-eclipsing spectroscopic binary can also allow a strong age constraint when present in a cluster with less evolved eclipsing systems.

V604 Lyr, also known as V56 \citep{Mochejska2002} and V96 \citep{Bruntt2003} (RA 19:20:45.26, DEC +37:45:48.6; \citealt{Gaia2018DR2}), is a photometric variable located slightly above the sub-giant branch (SGB) in the CMD of NGC\,6791, and we observed it with FLAMES at the Very Large Telescope for this reason. In this paper, we use the name V56, which we found to be a non-eclipsing SB2 binary with an orbital period of 10.822 days. Because of the location in an open cluster with well-measured eclipsing binaries \citep{Brogaard2011,Brogaard2012}, precise parameters of the sub-giant component of V56 can be extracted and used for constraining the cluster age, which is the aim of this paper. 

In Sect.~\ref{sec:data} we describe our spectroscopic observations.  Sect.~\ref{sec:stellarparms} explains how we extracted multi-epoch radial velocities of V56, the derivation of the spectroscopic orbit, and the determination of masses, effective temperatures, and radii through exploitation of the cluster CMD and the known properties of the eclipsing binaries V18 and V20 \citep{Brogaard2011,Brogaard2012}. We discuss the results in the context of the age of NGC6791 in Sect.~\ref{sec:results} and summarise our conclusions in Sect.~\ref{sec:conclusions}.

\section{Observations}
\label{sec:data}

We obtained spectra for a number of interesting stars in NGC\,6791 from FLAMES \citep{Pasquini2000} at the Very Large Telescope using the GIRAFFE spectrograph in Medusa mode (ESO programme ID 091.D-0125, PI: K. Brogaard). V56 was chosen as a target because it was a known photometric variable \citep{Mochejska2002,Bruntt2003}, although it was not known to be a binary prior to our observations. With the HR10 setting (533.9\,nm--561.9\,nm, central wavelength $548.8\, \mathrm{nm}$), the spectral resolution is $R=19800$, corresponding to 15.15 $\rm{km\,s^{-1}}$. We obtained 13 epochs of spectra, 11 with exposure times of 5400 seconds, the last two with an exposure time of 4800 seconds. The spectra were reduced using the standard instrument pipeline. For each target spectrum, a mean sky background from a number of sky fibres was subtracted.
Thorium-Argon (ThAr) wavelength calibration frames were not obtained during the nights of observation because the simultaneous ThAr option was disabled to avoid contamination in the spectra of the faint targets. Therefore, we applied radial velocity zero-point corrections to the spectra calibrated with standard day-time calibration frames. These corrections were calculated from changes in the nightly mean radial velocity of brighter giant stars that were also observed, as detailed in \citet{Brogaard2018}. After this procedure, the radial velocity epoch-to-epoch root mean square and half-range for single red giant branch members drop significantly from just above 0.5 $\rm{km\,s^{-1}}$ for both numbers to about 0.15 and 0.2 $\rm{km\,s^{-1}}$, respectively. We therefore estimate that the absolute uncertainty of the radial velocity zero-point of each observation is about $0.2\,\rm{km\,s^{-1}}$.

\section{Parameters of V56}
\label{sec:stellarparms}

We used the spectral information in combination with information and observations from our study of eclipsing binaries in NGC\,6791 to determine
the properties of V56 as detailed in this section.

\subsection{Radial velocity measurements}

For V56 we measured spectroscopic radial velocities (RVs) and a spectroscopic light ratio from the full available wavelength range in the single Giraffe order. This was done using a combination of the broadening function (BF) formalism \citep{Rucinski2002} and spectral separation \citep{Gonzalez2006}. Spectral lines from two components are visible in 12 epochs of the spectra making V56 a SB2 system. One epoch showed very significant RV overlap between components to a level that high precision RVs could not be extracted, even when using spectral separation. This spectrum was therefore left out of the analysis.
The measured RVs are given in Table~\ref{table:RV}. The uncertainties were first estimated by comparing RVs derived from the full spectral range to RVs derived from each half of the full range. This indicated uncertainties of the order 0.2\,km\,s$^{-1}$. However, to account for the RV zero-point uncertainty as well (cf. Sect.~\ref{sec:data}), we adopted total uncertainties of 0.3\,km\,s$^{-1}$, which is implied both by adding the two error contributions in quadrature, and the $\chi^2$  of our orbit solution, see Sect.~\ref{sec:orbit}.
Table~\ref{tab:EBdata} shows all our measurements of V56 based on the RVs and additional observations as presented in the following subsections.

\begin{table}
\centering
%\small
\caption{Radial velocity measurements of V56}
\label{table:RV}      % is used to refer this table in the text
    \begin{tabular}{lrr}
\hline
\hline
BJD-TDB & RV primary & RV secondary \\
 & $({\rm km}\cdot {\rm s}^{-1})$ & $({\rm km}\cdot {\rm s}^{-1})$ \\
\hline
56454.79039 & $-98.52\pm 0.30$ & $12.66\pm 0.30$ \\
56524.59051 & $5.17\pm 0.30$ & $-105.48\pm 0.30$ \\
56525.57976 & $0.32\pm 0.30$ & $-100.56\pm 0.30$ \\
56533.59017 & $-27.35\pm 0.30$ & $-68.71\pm 0.30$ \\
56540.52159 & $-96.91\pm 0.30$ & $10.31\pm 0.30$ \\
56477.75894 & $-77.34\pm 0.30$ & $-12.42\pm 0.30$ \\
56480.73664 & $0.38\pm 0.30$  & $-99.78\pm 0.30$ \\
56513.61523 & $4.62\pm 0.30$ & $-104.51\pm 0.30$ \\
56520.61633 & $-86.96\pm 0.30$ & $-1.38\pm 0.30$ \\
56517.63463 & $-73.72\pm 0.30$ & $-16.55\pm 0.30$ \\
56522.61280 & $-31.74\pm 0.30$ & $-63.56\pm 0.30$ \\
56523.57716 & $-7.22\pm 0.30$ & $-91.06\pm 0.30$ \\
\hline
    \end{tabular} 
\end{table}

\subsection{Spectroscopic orbit}
\label{sec:orbit}
We calculated orbit solutions of V56 with the RVs in Table~\ref{table:RV} using a Python implementation of the programme SBOP \citep{Etzel2004} created by one of us (J. Jessen-Hansen). When fitting the radial velocities, we allowed the systemic velocity of each component to vary independently because the components and their analysis could be affected differently by gravitational redshift and convective blueshift \citep{Gray2009} effects. As would be expected for a smaller and less evolved star from both effects, the derived values show that the secondary has a slightly more redshifted system velocity. Uncertainties on the derived system velocities are however too large to allow a meaningful detailed analysis of these physical effects. A solution forcing a common system velocity was also tried. That solution had systemic velocity $\gamma =-46.71$, slightly reduced semi-amplitudes, and thus smaller minimum masses by much less than $1\sigma$. The system velocity is fully consistent with cluster membership, since the mean cluster RV is $-47.40\pm0.13$\, km\,s$^{-1}$ with a $1\sigma$ velocity dispersion of $1.1\pm0.1$\, km\,s$^{-1}$ according to \citet{Tofflemire2014}.

The first solutions had eccentricity and argument of periastron as free parameters, but converged near zero eccentricity. Solutions with two system velocities had $e < 10^{-4}$ with an uncertainty larger than the eccentricity itself. When forcing one common system velocity we found $e=0.0047\pm0.0016$. This indicates that our measurements are not precise enough to either rule out or confirm a small eccentricity at a level of $5\cdot 10^{-3}$ or less. 
V56 would be expected to have circularised at the age of the cluster, as evidenced by the circularisation period in the similarly old open cluster NGC\,188 being 14.5 days according to \citet{Meibom2005}, and by comparing to the eclipsing binary V20 in NGC\,6791 which has a circular orbit with an orbital period of 14.47 days \citep{Brogaard2011}. However, the circularisation time scale is expected to be much longer than the synchronisation time scale, and we find indications in Sect.~\ref{sec:variability} that V56 is not synchronised, so perhaps it is not circularised either.
Whether true or not, we chose to fix the orbital solution to zero eccentricity, after verifying that this has negligible impact on the other parameters, especially the mass ratio $q$, which is the important number for reaching the absolute mass of the primary component of V56. 

The uncertainties in Table~\ref{table:RV} were estimated by increasing the systematic uncertainty from the radial velocity zero-point until our orbit solution returned a $\chi ^2$ of 1.0. The small increase needed of only $0.1\,\rm{km\,s^{-1}}$ suggests that the main uncertainty contribution is the RV zero-points, and thus all epochs and both components were estimated to have the same total uncertainty of $0.3\,\rm{km\,s^{-1}}$.  

\begin{figure}
	\includegraphics[width=9cm]{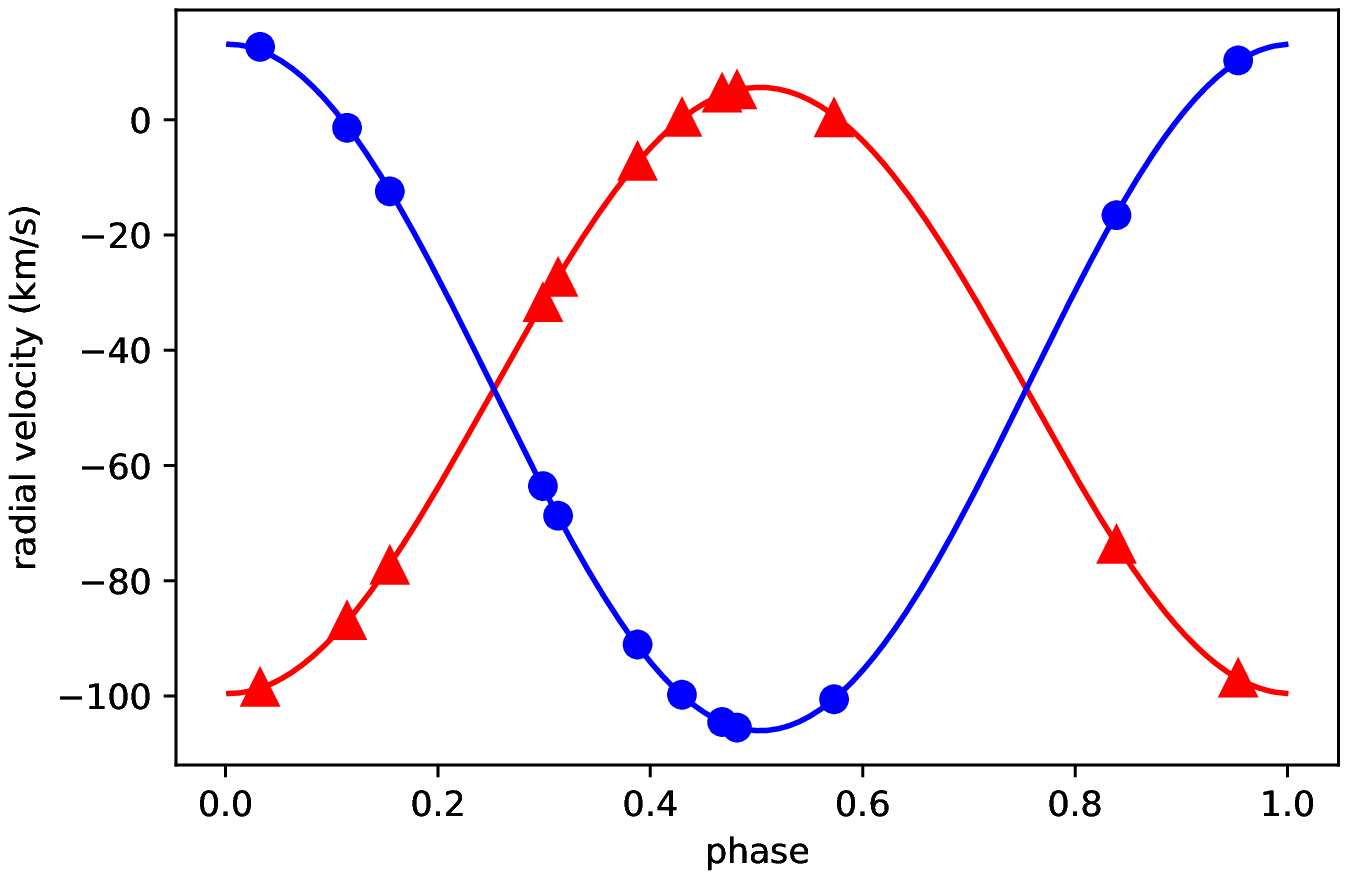}
	\includegraphics[width=9cm]{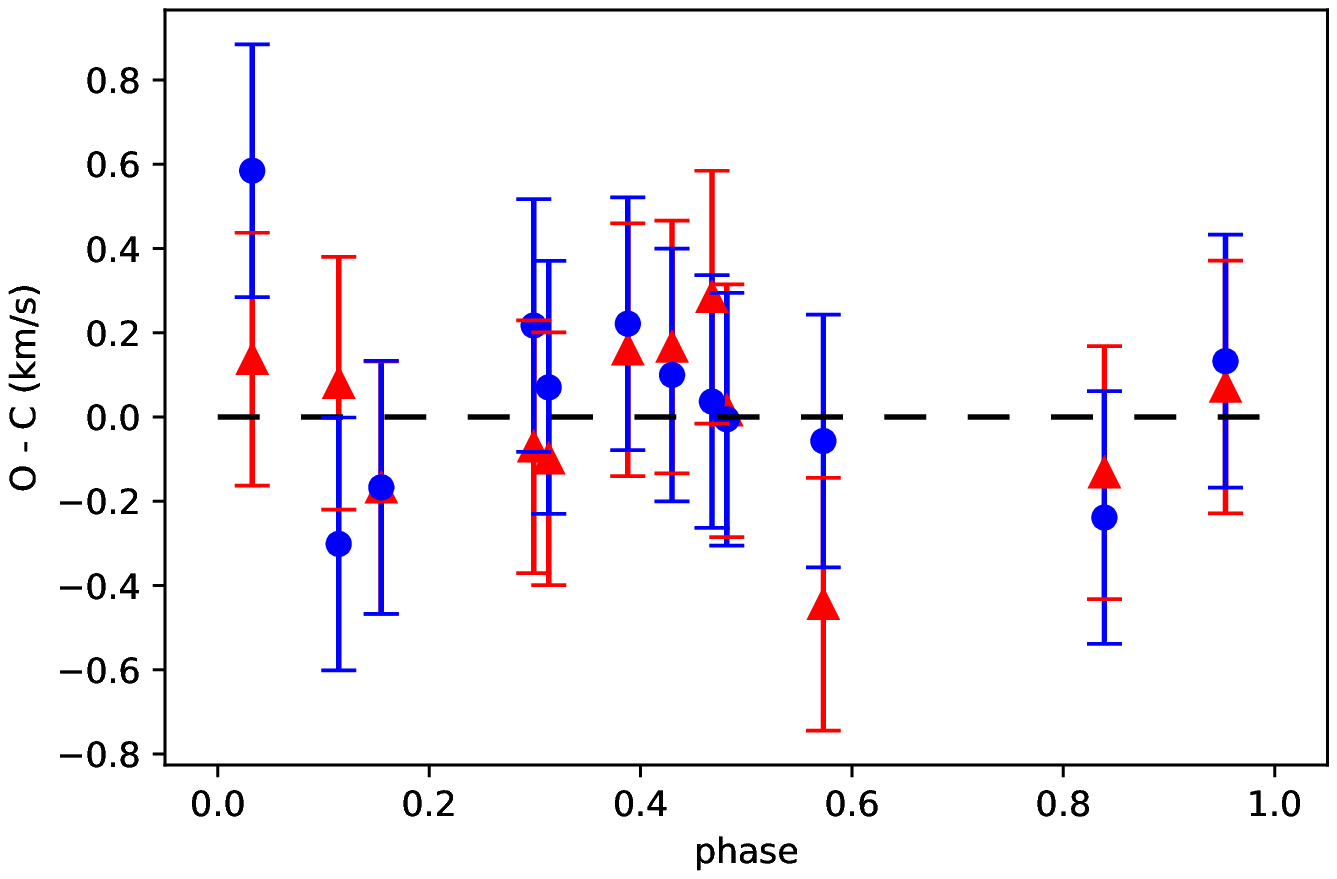}
	\caption{\textit{Upper panel:} SB2 orbit solution for V56. Red triangles are the RV measurements of the primary component and blue circles are the RV measurements of the secondary. \textit{Lower panel:} O-C diagram of the RV measurements relative to the model in the top panel.}
    \label{fig:orbit}
\end{figure}

Fig.~\ref{fig:orbit} shows the orbit solution and Table~\ref{tab:EBdata} displays the adopted spectroscopic parameters of V56, including the minimum masses. To obtain realistic uncertainties for the minimum masses, we recalculated the orbit as many times as we have epochs of observations, each time shifting the O-C pattern and recalculating. We estimated the uncertainty on the minimum masses from the minimum and maximum values obtained from these 12 solutions. The corresponding uncertainties for the semi-amplitudes are slightly larger than the uncertainty given by the adopted solution. 

\subsection{Light ratio}
\label{sec:lr}
The light ratio, $L_{\rm s}/L_{\rm p} = 0.386\pm0.020$, was determined as the ratio of the areas under the BF peaks of the separated spectra, calculated using separate templates \citep{Coelho2005} for each component with parameters selected by comparison to the eclipsing binary member V18 \citep{Brogaard2011,Brogaard2012} in the CMD. The uncertainty of $\pm0.020$, which is conservatively chosen to accommodate both random epoch-to-epoch scatter and a potential systematic bias due to differences in effective temperatures and log\,$g$ between components, was estimated by comparing to alternative, but less accurate, approaches: Using identical templates for both components resulted in $L_{\rm s}/L_{\rm p} = 0.399$. When calculating the light ratio directly in the observed spectra, using the procedure explained by \citet{Kaluzny2006}, which does not take into account the $T_{\rm eff}$ and log\,$g$ differences between components, we got $L_{\rm s}/L_{\rm p} = 0.41\pm0.006$, where the uncertainty is the standard deviation of the mean of individual epoch results. 

\subsection{Masses}

\begin{figure}
	\includegraphics[width=9.3cm]{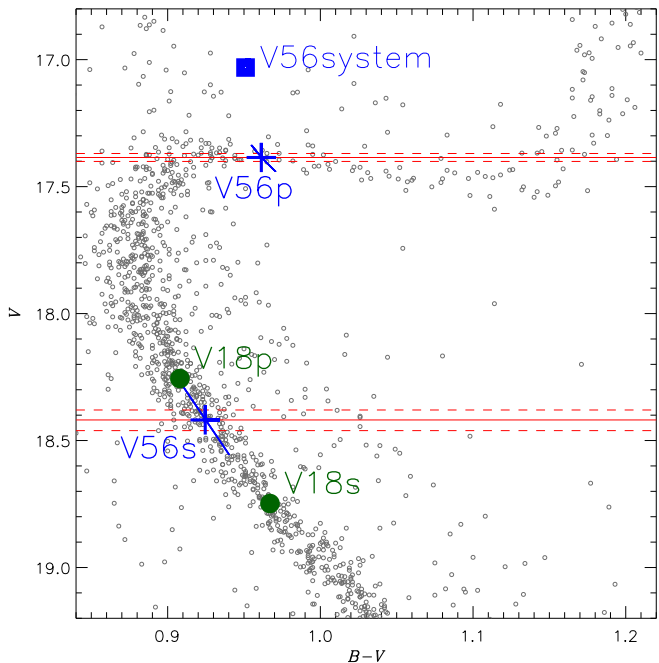}
	\caption{V56 in the CMD of NGC6791. Grey dots are the photometry of NGC\,6791 from \citet{Brogaard2012}. The blue square represents the total light of V56. The blue line segments correspond to a range of pairs of V56 component magnitudes that combine to the total light magnitude while having one component on the cluster main sequence. The solid red lines give the V56 component $V$-band magnitudes as calculated from the total light magnitude and the spectroscopic light ratio. Dashed red lines are the corresponding 1$\sigma$ uncertainties. Blue crosses are the best estimate V56 component magnitudes from the combined constraints of the cluster sequence and the spectroscopic light ratio.
	Green circles are the V18 component magnitudes.}
    \label{fig:V56}
\end{figure}

We measured the masses of the V56 components by combining the orbital solution with information from the cluster CMD and previous measurements of the eclipsing binary members V18 and V20 \citep{Brogaard2011,Brogaard2012}. By requiring that both components of V56 are located on the cluster fiducial sequence, as defined by a polynomial fit to the observed CMD, we found potential CMD positions for the individual components as . These positions, shown by the blue line-segments in Fig.~\ref{fig:V56}, were further constrained by making use of the light ratio $L_{\rm s}/L_{\rm p}=0.386\pm 0.020$ as calculated earlier from the spectra, cf. Sect.~\ref{sec:lr}. Adopting this light ratio constrained the $V$ magnitudes to regions marked by the red lines in Fig.~\ref{fig:V56}, and the best estimate of the CMD position of each component were chosen as the intersect of the red and blue line-segments. 
Since we know the mass at close-by locations on the CMD sequence from the components of the eclipsing member V18 (marked by green circles in Fig.~\ref{fig:V56}), we could interpolate linearly in mass along the sequence to obtain the mass at the position of the secondary of V56. We performed tests using isochrones to verify that linear interpolation in mass versus $V$-mag over the relevant interval is precise to a level of $0.0002$ $M_{\odot}$.
By comparing this mass of the V56 secondary to the minimum mass and the mass ratio found from the spectroscopic orbit, we obtained the orbit inclination and the mass of the primary component, which is on the subgiant branch. The masses of V56 calculated in this way turn out to be $1.103\pm 0.006 M_{\odot}$ and $0.974\pm 0.006 M_{\odot}$ for the primary and secondary components, respectively. 
To these uncertainties, we add for both stars in quadrature the contribution from the uncertainty on the masses of the V18 components, $\pm0.0033$ $M_{\odot}$ \citep{Brogaard2012}. The uncertainty on the V56 primary mass includes the additional error contribution from the uncertainty on the mass ratio $q$ of V56 from the orbit solution, which amounts to $\pm0.0040$ $M_{\odot}$, which was also added in quadrature. Including all the error contributions, our mass estimates for V56 are $M_p=1.103\pm 0.008 M_{\odot}$ and $M_s=0.974\pm 0.007 M_{\odot}$.

\subsection{Photometric variability}
\label{sec:variability}

\begin{figure}
	\includegraphics[width=9.5cm]{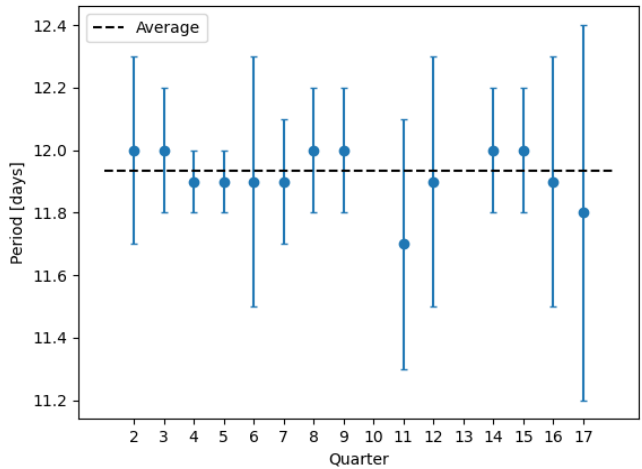}
	\caption{Photometric periods of individual \textit{Kepler} quarters determined by-eye. Weighted average is $11.93\pm0.02$ days.}
    \label{fig:periods}
\end{figure}

\begin{figure}
	\includegraphics[width=8.4cm]{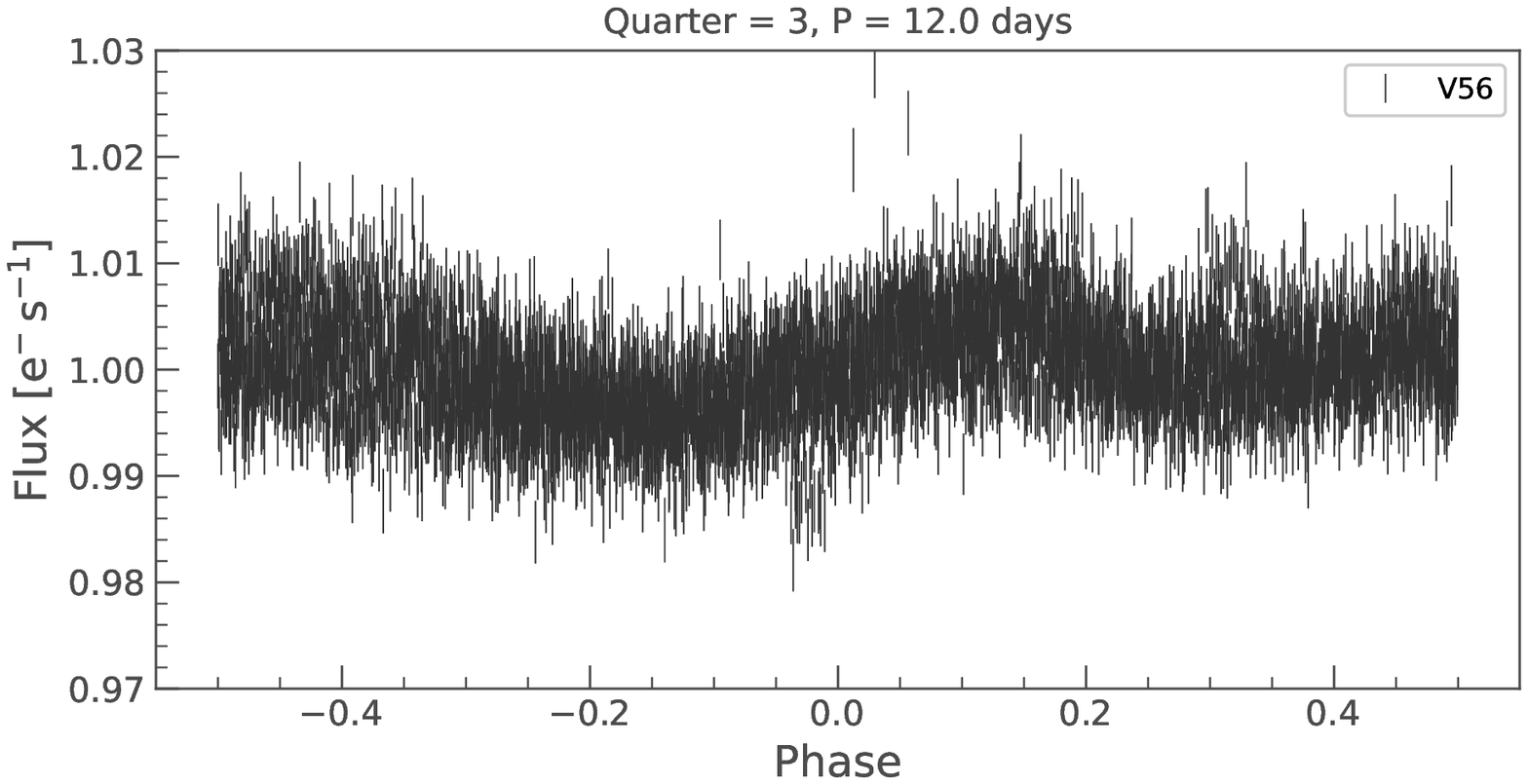}    
	\includegraphics[width=8.4cm]{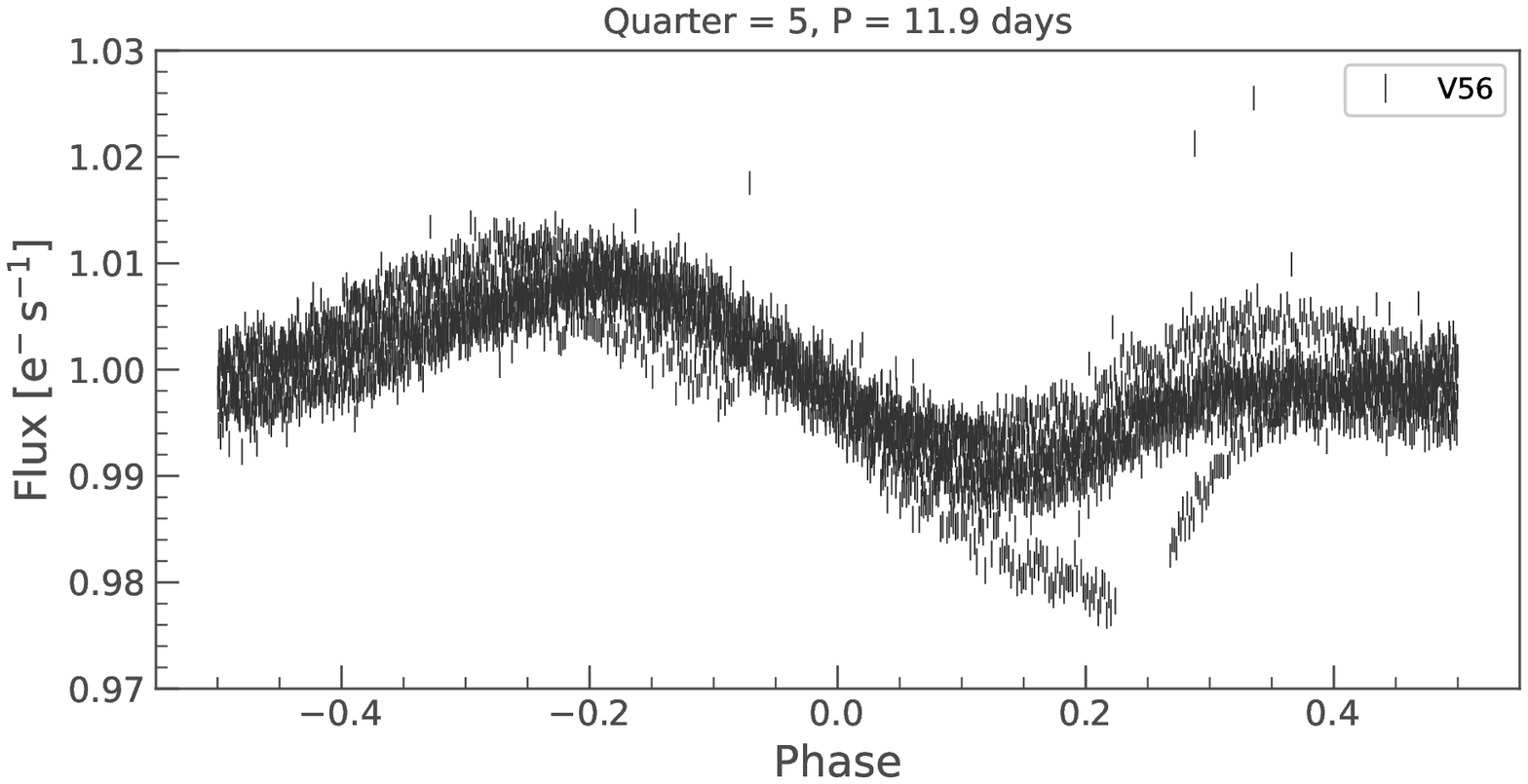}
	\includegraphics[width=8.4cm]{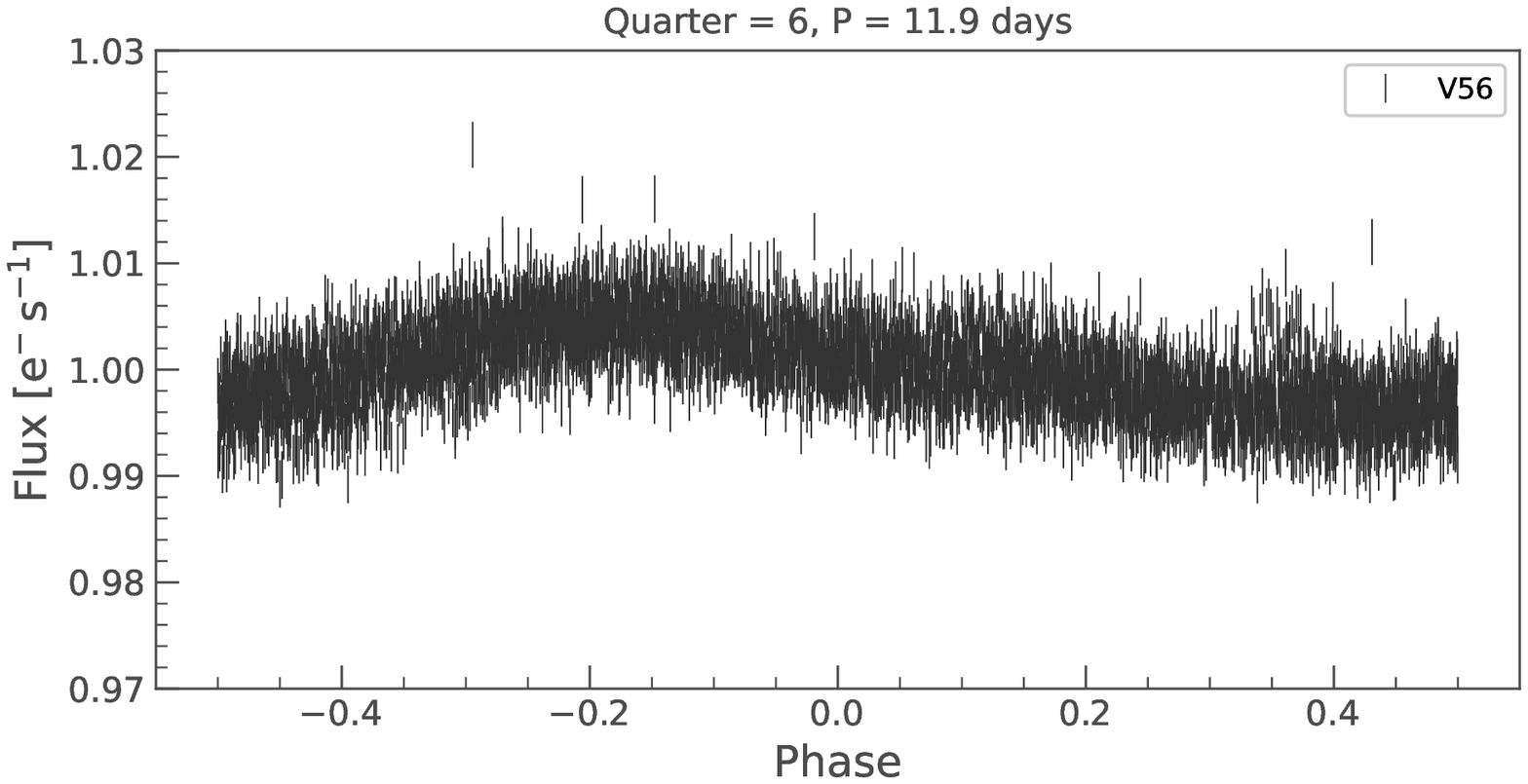}
	\includegraphics[width=8.4cm]{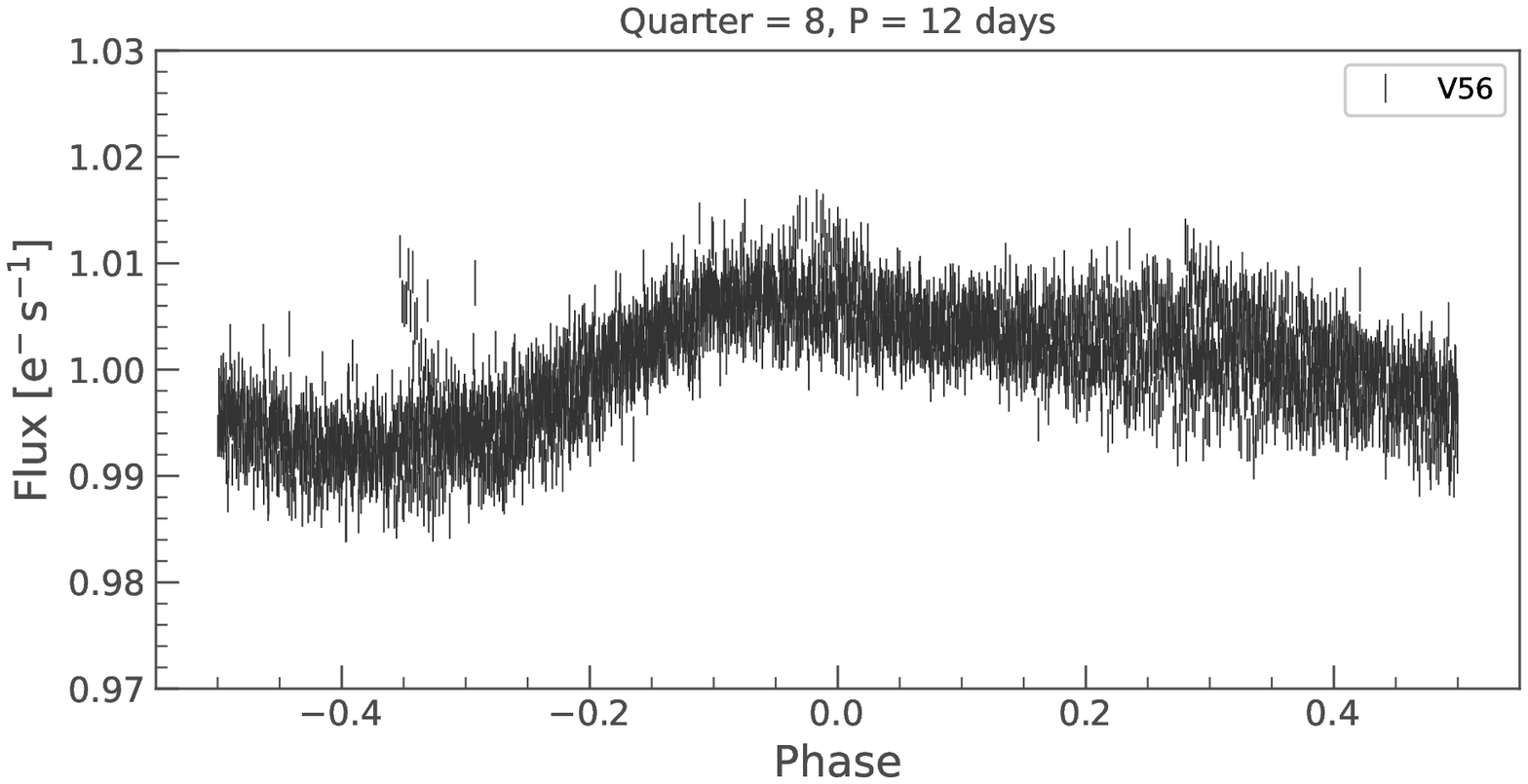}
	\includegraphics[width=8.4cm]{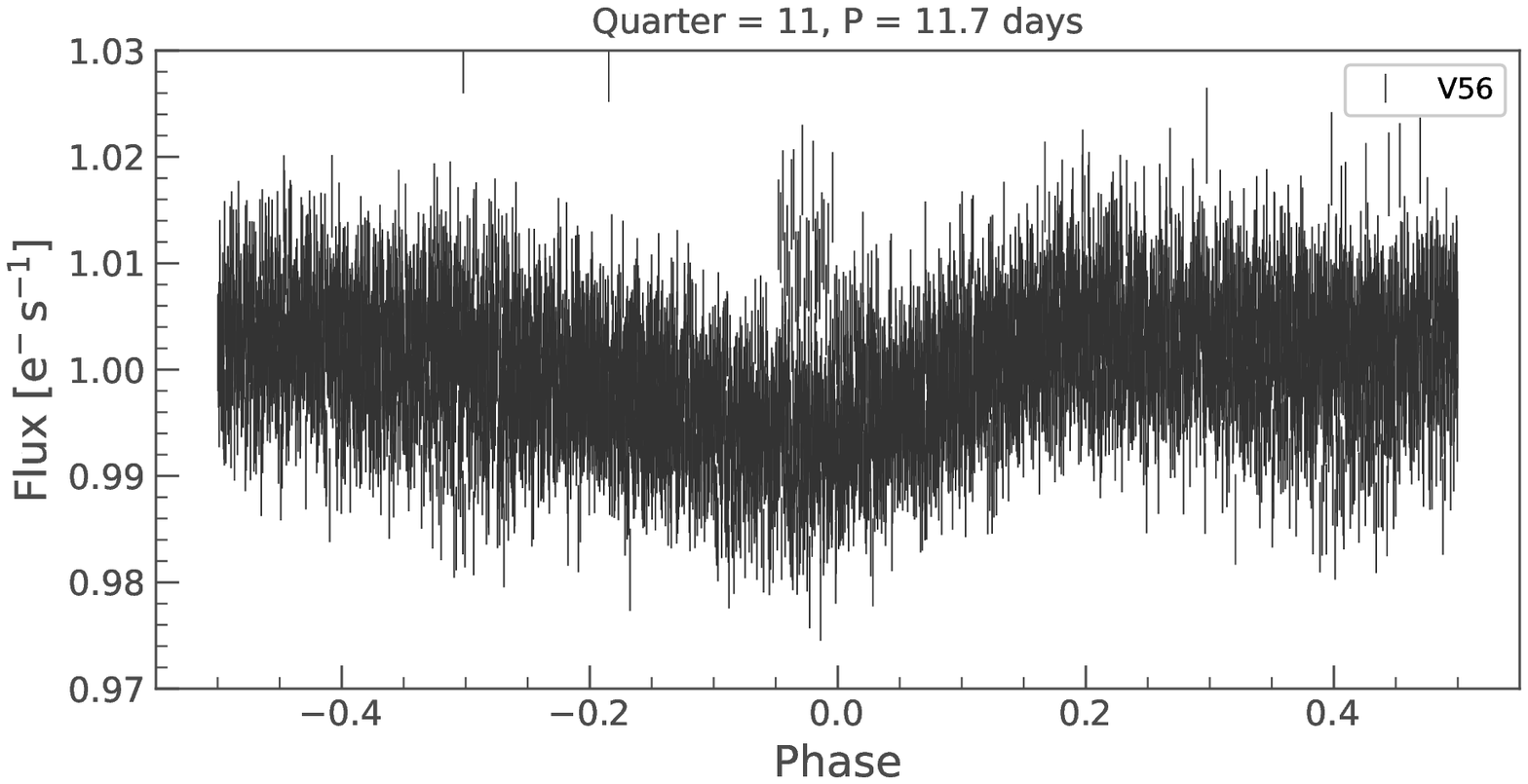}
	\includegraphics[width=8.4cm]{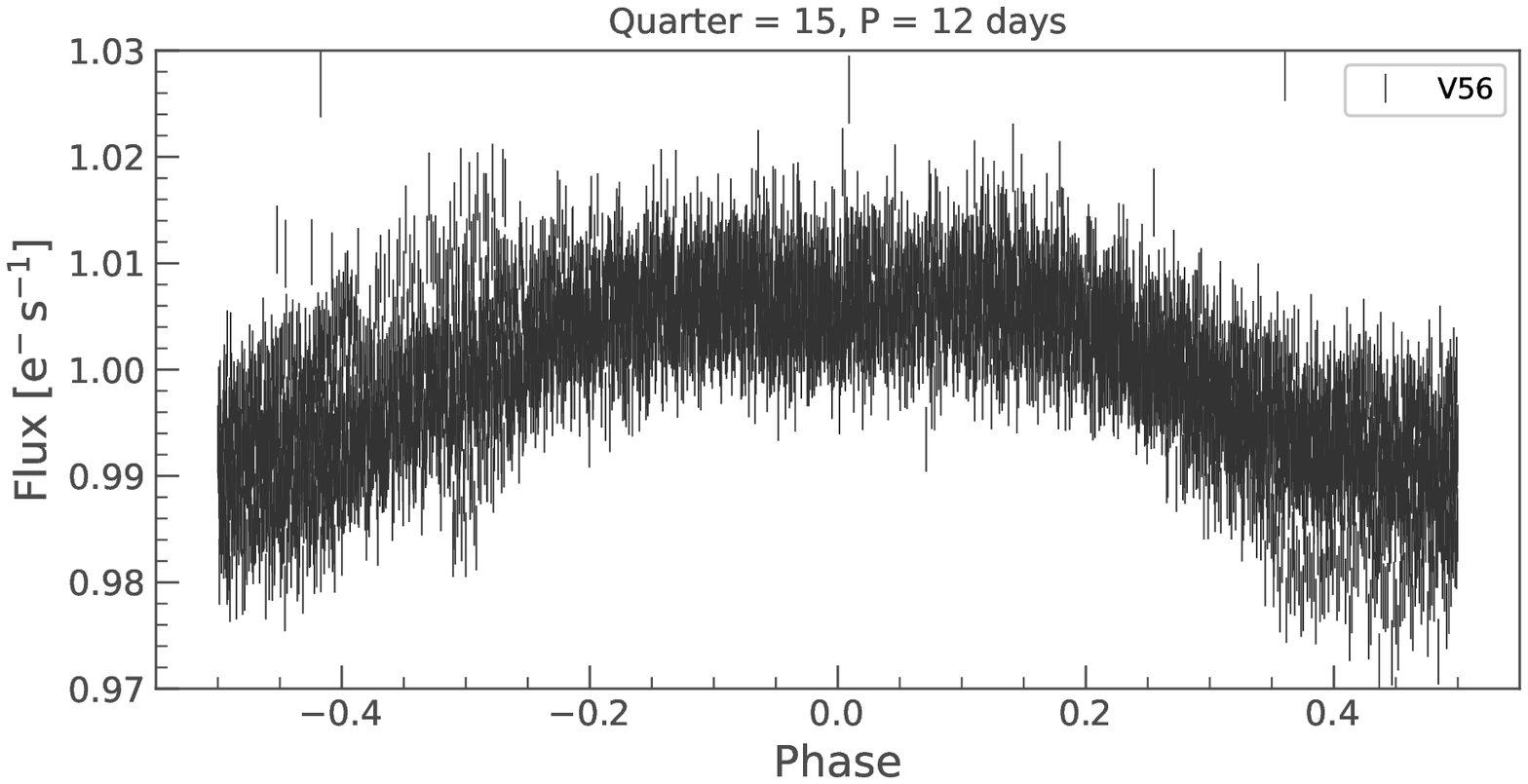}
	\caption{\textit{Kepler} light curves of V56 for selected quarters folded with the best estimate of the photometric period.}
    \label{fig:lightcurves}
\end{figure}

Since V56 is a known photometric variable but of unknown type, we investigated the {\it Kepler} light curve that we extracted from the so-called superstamps of NGC\,6791 \citep{Kuehn2015}. 
We downloaded the superstamps for all 17 quarters from MAST. For each quarter we created a one-pixel mask at the coordinate location of V56 to obtain single-pixel light curves using the Python program Lightkurve \citep{Lightkurve}. Quarters 1,10, and 13 did not produce meaningful light curves due to bad pixels and/or contaminating neighbouring stars. From the other light curves we created phase plots adjusting the period manually to obtain the smallest scatter by eye. The periods so obtained are shown in Fig.~\ref{fig:periods} and the weighted mean of these is $11.93\pm0.02$ days. Alternatively, we also determined the period using the periodogram function in Lightkurve, which resulted in a mean period of 12.26 days. The light curves for selected individual quarters are shown in Fig.~\ref{fig:lightcurves}. The photometric period is clearly different from the orbital period, and the amplitude and shape of the variability changes with time. We therefore attribute the variability to spots. The difference between the orbital period, 10.82 days, and the photometric period, 11.93 days, will then have to be attributed to a combination of differential rotation, spot evolution and potential lack of synchronisation. The robustness of the photometric period over long time scales and the relatively large difference between orbit and photometric period seem to suggest the orbit has not synchronised. This is at odds with theory. According to \citet{Zahn1977}, synchronisation should happen much faster than circularisation, and in the case of V56 the time scale $t_{\rm sync}$ is less than 0.2 Gyr. Thus something seems to be, or to have been, disturbing V56 and preventing synchronisation.

As in the ground based light curves by \citealt{Bruntt2003} where the star has ID V96 the peak-to-peak variability amplitude never exceeds 0.024 mag, except for observational uncertainty and outliers. Since the ground based $B$ and $V$ magnitudes are calculated as the mean of many measurements taken on many different nights we estimate the maximal error in the mean magnitude and $B-V$ colour of V56 is $0.01$ mag. We checked that such a change is inconsequential for our conclusions regarding the mass estimates which remain well within the stated uncertainties (changes are of the order $\pm0.002 M_{\odot}$ or less), but the exact colour-location of the primary component on the sub-giant branch can be different by as much as $\pm0.015$ mag in $B-V$ colour. Potential differential reddening effects \citep{Platais2011,Brogaard2012} would also cause effects of this order of magnitude. We therefore conclude that we are able to predict the masses of the components of V56 with a precision better than 1\% but that the exact CMD position of the sub-giant component has a colour-uncertainty in $B-V$ of about $\pm0.02$ mag.

\begin{table}
%\centering
%\small
\caption{Parameters of V56.}
\label{tab:EBdata}      % is used to refer this table in the text
    \begin{tabular}{lr}
\hline
\hline
Parameter & Value \\
\hline
phot ID\tablefootmark{$^1$} & 15162 \\
$V_{\rm tot}$ & 17.031(15) \\
$V_{\rm p}$ & 17.385(16) \\
$V_{\rm s}$ & 18.419(42) \\
$B_{\rm tot}$ &  17.982(15)\\
$B_{\rm p}$ & 18.347(18)\\
$B_{\rm s}$ & 19.344(46) \\
$(B-V)_{\rm tot}$ & 0.951(21)\\
$(B-V)_{\rm p}$ & 0.961(24)\\
$(B-V)_{\rm s}$ & 0.925(62)\\
\hline
$P$ (days) & 10.8219(17)   \\
$e$ & 0\tablefootmark{$^2$} \\
$K_{\rm p} ({\rm km}\,{\rm s}^{-1})$ & 52.54(10) \\
$K_{\rm s} ({\rm km}\,{\rm s}^{-1})$ & 59.49(11) \\
$q$ & $0.8829(30)$ \\
$\gamma_{\rm p} ({\rm km}\,{\rm s}^{-1})$ & -46.80(8) \\
$\gamma_{\rm s} ({\rm km}\,{\rm s}^{-1})$ & -46.59(9)  \\
$M_{\rm p}{\rm sin^{3}}i \rm(M_{\odot})$ & $0.8372_{-0.0073}^{+0.0021}$         \\
$M_{\rm s}{\rm sin^{3}}i \rm(M_{\odot})$ & $0.7394_{-0.0032}^{+0.0031}$    \\
$L_{\rm s}/L_{\rm p}$ & $0.386\pm0.020$  \\
\hline
$M_{\rm p} \rm(M_{\odot})$ & $1.103\pm 0.008$ \\
$M_{\rm s} \rm(M_{\odot})$ & $0.974\pm 0.007$ \\
$R_{\rm p}({\rm R_{\odot}})$ & $1.764\pm0.099$ \\ 
$R_{\rm s}({\rm R_{\odot}})$ & $1.045\pm0.057$ \\
$T_{\rm eff,p}$ (K) & $5447\pm125$ \\
$T_{\rm eff,s}$ (K) &$ 5552\pm125$ \\
$\rm{[Fe/H]}$ & $+0.29\pm0.06$ \\
$i (^{\circ})$ & $65.76\pm0.13$\\
\hline
    \end{tabular} 
        \tablefoot{\\
        \tablefoottext{$^1$}{Photometry from \citet{Brogaard2012}\\}
\tablefoottext{$^2$}{The value was fixed.} 
}
\end{table}

\subsection{$T_{\rm eff}$}
\label{sec:teff}
The CMD positions of the V56 components obtained above yields also the component $(B-V)$ colours that can lead to $T_{\rm eff}$ through a colour-$T_{\rm eff}$ relation. To minimise the effect of uncertainty in the reddening, we adjusted $E(B-V)$ in the bolometric correction calculations of \citet{Casagrande2014} until the colours of the components of V18 were matched at their spectroscopic $T_{\rm eff}$ values by \citet{Brogaard2012}. We then used this $E(B-V)$ value to obtain the $T_{\rm eff}$ values corresponding to the colours of the V56 components. This yielded $T_{\rm eff,p}=5447\pm125$\,K and $T_{\rm eff,s}=5552\pm125$\,K, with the uncertainties conservatively adopted from the largest of the uncertainties on the spectroscopic $T_{\rm eff}$ values for the V18 components. We adopted $[\rm{ Fe/H}]=+0.35$ \citep{Brogaard2012}, but using instead $[\rm{ Fe/H}]=+0.29$ caused changes of only a few K because it is a relative measurement. 

\subsection{Radii}

The spectral resolution of our observations was $R$ = 19800, corresponding to 15 ${\rm km}\,{\rm s}^{-1}$. The expected rotational velocities $v\sin (i)$ of V56 are significantly smaller. This made us unable to calculate the component radii from $v\sin(i)$ measurements as done for the much faster rotating V106 by \citet[their eqn. 1]{Brogaard2018}. Our V56 radii determined as explained below predict $v\sin (i)$ values of 6.8-7.5 ${\rm km}\,{\rm s}^{-1}$ and 4.0-4.5 ${\rm km}\,{\rm s}^{-1}$ for the primary and secondary, respectively, depending on whether the orbit period or the photometric period (cf. Sect.\ref{sec:variability}) is adopted. Because these are only a fraction of the spectral resolution, and because the spectral lines are further broadened an additional 1-2 ${\rm km}\,{\rm s}^{-1}$ by the change in radial velocities during the 5400 second long exposures, we were not able to measure meaningful $v\sin (i)$ values for V56. Instead, we adopted the below procedure to estimate the radii of the V56 components. 

We first used equation (10) of \citet{Torres2010} to calculate the absolute component magnitudes:
\begin{equation}
M_V=-2.5\rm{log}\left(\frac{\it L}{{\it L}_\odot}\right)+{\it V}_\odot+31.572-\left(BC_{\it{V}}-BC_{\it{V,\odot}}\right).
\end{equation}

Here, $V_\odot=-26.76$ as recommended by \cite{Torres2010} and BC$_{V,\odot}=-0.068$ as obtained from the calibration of \citet{Casagrande2014}, which was also used to obtain the bolometric corrections.

We then combined equation (1) with $L=4\pi \sigma R^2T_{\rm eff}^4$, $\sigma$ being the Stefan-Boltzmann constant, and the apparent distance modulus to calculate the radius:

\begin{equation}
\frac{R}{R_{\odot}}=\left(\frac{T_{\rm eff}}{T_{\rm eff,\odot}}\right)^{-2}\cdot 10^{\frac{(V-M_V)-V+V_{\odot}+31.572-\rm{BC}_{\it{V}}+\rm{BC}_{\it{V,\odot}}}{5}}.
\end{equation}

The $V$-band was utilised because for this filter we have both a spectroscopic light ratio for V56 and a measurement of the apparent distance modulus $(V-M_V)=13.51$ from \citet{Brogaard2012}. With these, eqn. (2) yields $R_p = 1.764\pm0.099 R_{\odot}$ and $R_s=1.045\pm0.057 R_{\odot}$. We adopted a 125K uncertainty on $T_{\rm eff}$ with corresponding bolometric corrections assuming $[{\rm Fe/H}]=+0.35$. An additional error contribution of $\pm0.005 {R_{\odot}}$ for the primary and $\pm0.002 {R_{\odot}}$ arises from a change in $\rm [Fe/H]$ of $\pm0.06$. If V18 and V56 are not affected in the same way by small scale differential reddening \citep{Platais2011,Brogaard2012} that might add an additional small effect. 
We have not considered an uncertainty on the apparent distance modulus because this number comes from the analysis of V18 (and V20), and is therefore strongly correlated with the effective temperatures for V18, which are already the sources for our adopted $T_{\rm eff}$ uncertainties.

\subsection{Atmospheric parameters}
\label{sec:atmosphere}
\begin{figure*}
	\includegraphics[]{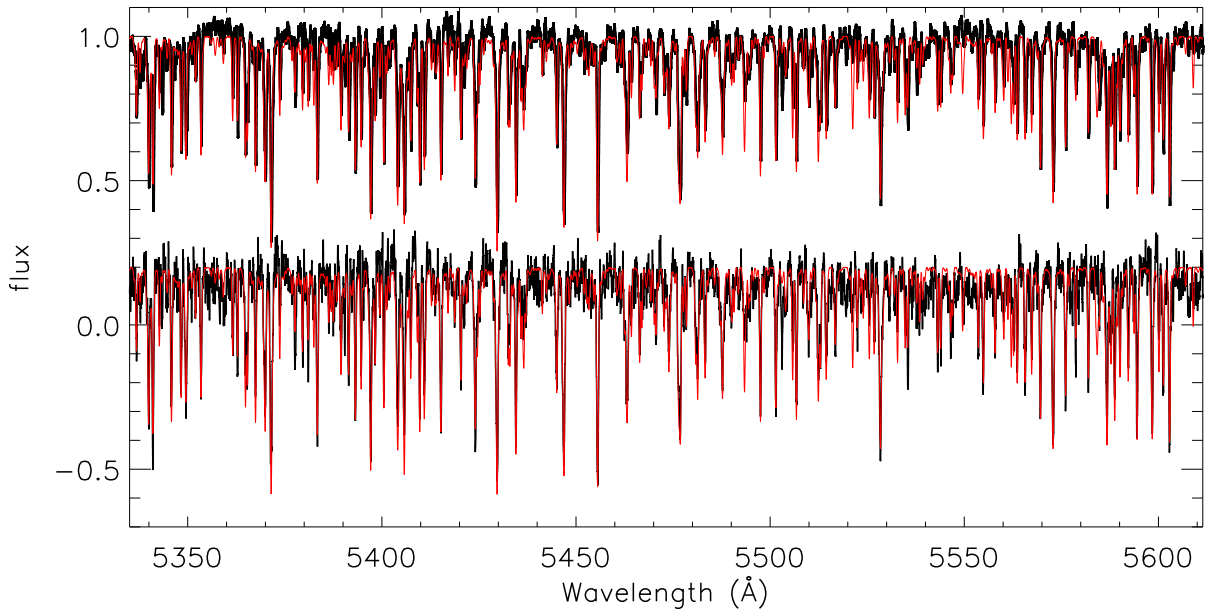}
	\caption{Disentangled and renormalised spectra of the V56 components, primary on top, secondary below. The flux level of the secondary spectrum has been shifted from by -0.8 from 1 for clarity. The observed spectra are shown in black, the model spectra from Sect.~\ref{sec:atmosphere} are overplotted in red.}
    \label{fig:spectra}
\end{figure*}

We carried out a classical equivalent-width spectral analysis on the disentangled spectra to determine $T_{\rm eff}$ and $\rm [Fe/H]$ for the two components following the method outlined in \citet{Slumstrup2019}. We fixed $\log g$ to the values from the binary solution, which are 3.99 and 4.39 for the primary and secondary, respectively. This was determined using the masses and radii from Table~\ref{tab:EBdata} with $\log g = \log ( {GM}/{R^2} ) $ in solar units. The atmospheric parameters were determined with the auxiliary program Abundance with SPECTRUM \citep{Gray1994} assuming LTE, using MARCS stellar atmosphere models \citep{Gustafsson2008} and solar abundances from \citet{Grevesse2007}. This yielded $T_{\rm eff}=5490 \pm 150$\,K, $\rm [Fe/H]=0.27 \pm 0.07$ for the primary and $T_{\rm eff}=5550 \pm 190$\,K, $\rm [Fe/H]=0.34 \pm 0.10$ for the secondary. The uncertainties are only internal and do not include systematic effects. We show the renormalised disentangled spectra and the corresponding best matching theoretical spectra in Fig.~\ref{fig:spectra}\footnote{The disentangled spectra are available in tabular form as Tables A1 and A2 at the CDS.}. 

Comparison of these $T_{\rm eff}$ values to those derived from photometry in Sect.~\ref{sec:teff} show excellent agreement with differences of only 43 K and 2 K for the primary and secondary, respectively, much less than the $1\sigma$ uncertainties. The [Fe/H] values derived are also in excellent agreement with those of \citet{Brogaard2011}. The self-consistency of both $T_{\rm eff}$ and [Fe/H] values provide further support for our procedure and derived parameters. We adopt the weighted mean of [Fe/H] of the two V56 components, $+0.29\pm0.06$, as the metallicity of V56 in Table~\ref{tab:EBdata} while noting that this does not include any systematic uncertainty, and that this is the atmospheric abundance.

\section{Results \& discussion}
\label{sec:results}

\begin{figure*}
	\includegraphics[width=18cm]{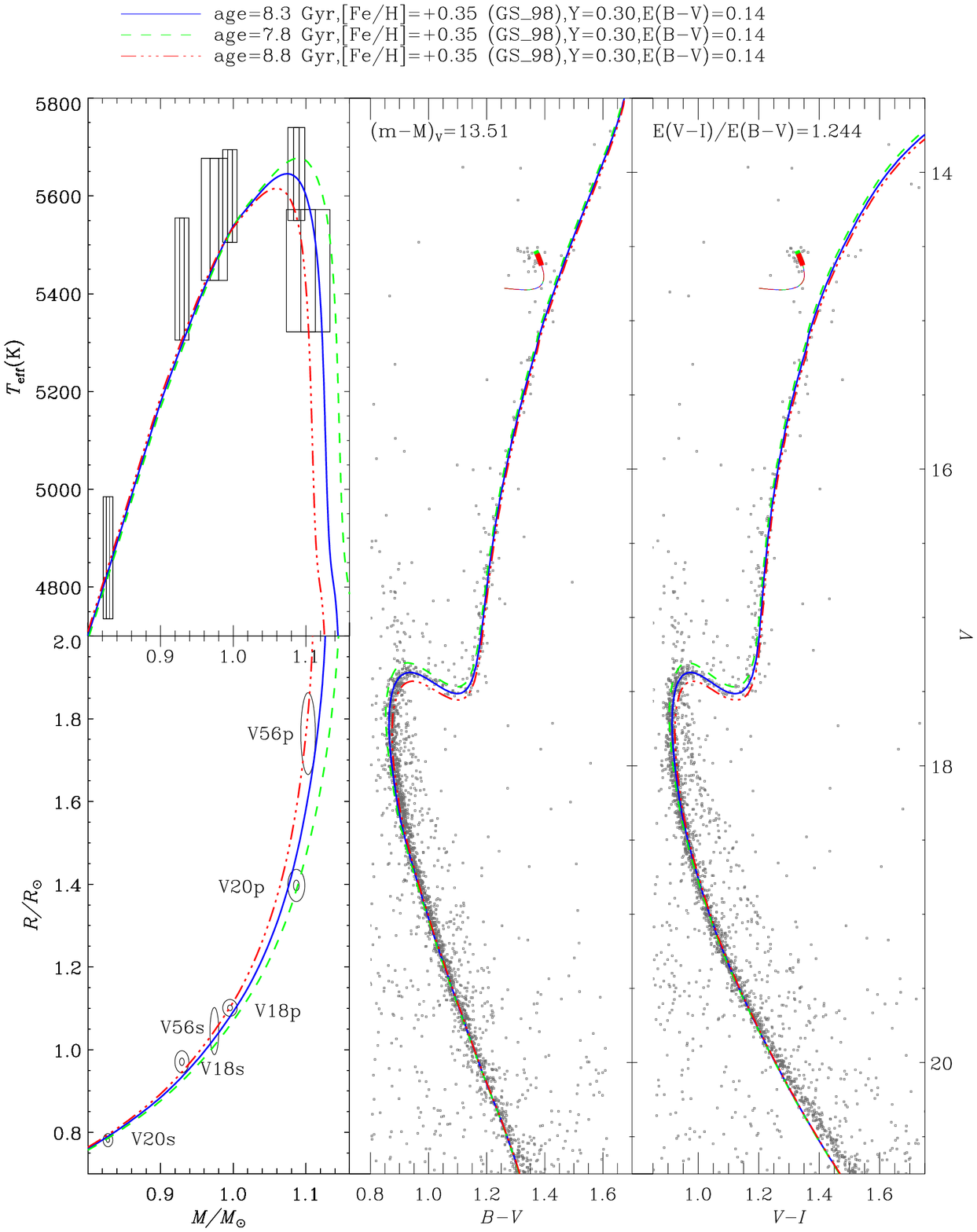}
	\caption{Measurements of V56 and the eclipsing binaries V18 and V20 (left panels) and cluster CMDs (middle and right panels) compared to Victoria
model isochrones and ZAHB loci as described in \citet{Brogaard2012}. Uncertainties shown in the left panels are 1$\sigma$ and 3$\sigma$ for mass and radius of V18 and V20, but only 1$\sigma$ for V56, and 1$\sigma$ for all $T_{\rm eff}$ values. GS\_98 denotes the solar abundance pattern of \citet{grevesse1998}. The thick part of the ZAHB corresponds to masses with an RGB mass-loss from zero to twice the mean mass loss found from asteroseismology by \citet{Miglio2012}.}
    \label{fig:model}
\end{figure*}

We show our measurements in Fig.~\ref{fig:model}, which is a reproduction of fig. 4 from \citet{Brogaard2012}, now expanded to include V56 in the mass-radius and mass-$T_{\rm eff}$ diagrams.
As seen from the mass-radius diagram, the measurements of V56 are consistent within 1$\sigma$ with the best solution found from V18 and V20. Unsurprisingly, the secondary component of V56 does not constrain the isochrones additionally, because its properties were inferred from the very similar stars in V18 that bracket the V56 secondary. The primary component is however significantly more evolved than any of the other stars in the mass-radius diagram, and therefore provides constraints for the isochrones. The larger measurement uncertainties compared to the eclipsing systems are partly compensated by larger differences between the isochrones at the larger mass corresponding to more evolved stars. The exact best estimate values for mass and radius of the V56 primary suggest an age older by only 0.4 Gyr compared to that of the combined analysis of V18 and V20 from \citet{Brogaard2012} while being consistent with that solution within 1$\sigma$.
Our analysis of V56 therefore supports the age estimates and conclusions of \citet{Brogaard2012} for NGC\,6791. Furthermore, it emphasises the potential of analysing similar non-eclipsing SB2 systems in other star clusters with known eclipsing binary systems that are not sufficiently evolved to constrain the cluster age.

As seen in Fig.~\ref{fig:model}, an age close to 8.3 Gyr best reproduces the mass-radius relation as determined by V18, V20 and V56 together with the cluster CMD. In the past, age estimates of NGC\,6791 varied quite a lot from one investigation to another, but with the mass-radius and mass-$T_{\rm eff}$ relations constrained as they are now, the wiggle-room on the age has decreased significantly. This has implications for other works on, for example, the age of the cluster and on attempts to empirically calibrate asteroseismic scaling relations. Yet, surprisingly, there are still age estimates in recent literature that deviate quite a lot as detailed below.

\subsection{Age in cluster catalogues}

\citet{Kharchenko2013} determined log(age) = 9.645 for NGC\,6791 corresponding to an age of only 4.42 Gyr. We inspected the details and found two issues that explain this as a systematic error caused by the facts that they assume solar metallicity and, most importantly, the photometry that they use does not reach the cluster turn-off and main sequence. This also explains why the distance that they determine along with the age, $d = 4926\,pc$, was not problematic in their analysis: They were not limited by having to match the turn-off and main-sequence at all. 

\citet{Cantat2020} used a set of open clusters with well-determined parameters in the literature to train an artificial neural network to estimate parameters (ages, extinctions, and distances) from Gaia photometry and parallax. For NGC\,6791 they determined log(age)=9.80, which translates to 6.31 Gyr. While being closer to our estimate than that of \citet{Kharchenko2013}, it is still too young to seem trustworthy. The relatively low parallax value (=large distance) and especially the very large extinction value $A_V=0.7$ that they determined along with the young age enables a reasonable isochrone match to most of the CMD with their parameters, but the mass-radius relation would then be completely off for the binaries, the reddening much larger than any previous determination, and the observed red clump stars would be significantly brighter than the model. As mentioned by \citet{Cantat2020}, their procedure does not account for variations in metallicity. We therefore suspect that their large extinction, which in turn causes their young age, is an artefact of the neural network trying to account for the high super-solar metallicity of NGC\,6791 by increasing reddening and thus extinction. 

\subsection{Age and asteroseimic scaling relations}

\citet{Kallinger2018} derived an age of $10.1\pm0.9$ Gyr and true distance modulus of $(m-M)_0=13.11\pm0.03$ for NGC\,6791 based on asteroseismology of red giants along with empirical corrections to asteroseismic scaling relations. Their empirical corrections to the scaling relations were derived by forcing agreement between literature dynamical parameters \citep{Gaulme2016,Brogaard2018A,Themessl2018} and their own asteroseismic measurements for a sample of selected eclipsing binaries with an oscillating giant component. This requires that the dynamical parameters of eclipsing binary stars are reliable. However, their fig. 11 shows that the age they derive for NGC\,6791 is not consistent with the properties of the primary stars of the eclipsing members V18, V80 and V20. The disagreement is rather large; the x-axis is logarithmic and the mass of the V20 primary component is erroneously shown at a lower value rather than the true measurement of \citet{Brogaard2011} of $1.0868\pm0.0039 M_{\odot}$. With the addition of the V56 primary properties from the present work, the discrepancy is now even more obvious. Their derived distance modulus for NGC\,6791 also works against them: They assumed $E(B-V)=0.15$ (T. Kallinger, priv. comm) which yields $(m-M)_V=13.575$ adopting a standard $A_V=3.1\times E(B-V)$. Comparing that to Fig.~\ref{fig:model}, this requires a slightly younger age than 8.3 Gyr, and is not consistent with their much older age close to 10 Gyr.  
Since there is no reason to put more reliance on the dynamical parameters of the binary stars used to derive the empirical corrections to the scaling relations than those of the binary stars in the cluster or even the cluster CMD, one must question the accuracy of both the age of NGC\,6791 and scaling relation corrections derived by \citet{Kallinger2018}. The cause of the problem is likely that the calibration sample used by \citet{Kallinger2018} is too small, which they also caution.

\citet{Brogaard2012} gave an estimate of the mass of the RGB stars at the $V$-band luminosity of the red clump in NGC\,6791, $M_{\rm RGB,NGC\,6791}=1.15\pm0.02 M_{\odot}$ and made a first comparison to early asteroseismic measures, showing that the asteroseismic scaling relations, which at that time were used without corrections, overpredicted the mass of RGB stars in NGC\,6791 \citep{Basu2011,Miglio2012}. Those measurements, supported also by our new measurements of V56, now show that the empirical corrections suggested by \citet{Kallinger2018} make the scaling relations underpredict mass, since they suggest that the cluster RGB mass below the RC, 1.10 $M_{\odot}$ as measured by \citet{Kallinger2018}, is lower than the early SGB mass, 1.103 $M_{\odot}$ as represented by V56p, at odds with stellar evolution theory. Theoretically predicted corrections to the asteroseismic scaling relations such as those of \citet{Rodrigues2017} are smaller than suggested empirically by \citet{Kallinger2018} and could lead to self-consistent results for the giant stars of NGC\,6791, but their accuracy is still uncertain. Future attempts to further test and/or calibrate asteroseismology using NGC\,6791 \citep{Sharma2016,Pinsonneault2018} and additional open clusters and eclipsing binaries are therefore still needed.

\section{Conclusions}
\label{sec:conclusions}

We identified V56 as a non-eclipsing spectroscopic binary member of the open cluster NGC\,6791 with two visible spectral components. Multi-epoch spectra were exploited in combination with prior knowledge and observations of NGC\,6791 and its known eclipsing members to obtain masses, radii, $T_{\rm eff}$ values and [Fe/H] for V56. These properties were then compared to isochrones, in combination with those of the eclipsing members V18 and V20 and the cluster CMD. This comparison support and strengthen the conclusions of \citet{Brogaard2012} with respect to a cluster age of $8.3\pm0.3$ Gyr and a mass of the RGB stars of $M_{\rm RGB,NGC\,6791}=1.15\pm0.02 M_{\odot}$. These numbers therefore continue to serve as verification points for other age dating methods, such as various inferences from cluster CMDs and mass measures, for example in asteroseismology. We encourage the search for, and identification and exploitation of similar systems to V56 in other open clusters with known eclipsing binaries.

\section*{Acknowledgements}

We thank D.\,A.\,VandenBerg for assistance in the early stages of the project and for providing the stellar models used in Fig.~\ref{fig:model}.\\
We thank the anonymous referee for useful comments.\\
Based on observations collected at the European Southern Observatory under ESO programme 091.D-0125(A).\\
We gratefully acknowledge the grant from the European Social Fund via the Lithuanian Science Council (LMTLT) grant No. 09.3.3-LMT-K-712-01-0103.\\
Funding for the Stellar Astrophysics Centre is provided by The Danish National Research Foundation (Grant agreement no.: DNRF106).\\
This research has made use of the SIMBAD database,
operated at CDS, Strasbourg, France\\
This paper includes data collected by the Kepler mission and obtained from the MAST data archive at the Space Telescope Science Institute (STScI). Funding for the Kepler mission is provided by the NASA Science Mission Directorate. STScI is operated by the Association of Universities for Research in Astronomy, Inc., under NASA contract NAS 5–26555.

   \bibliographystyle{aa} % style aa.bst
   \bibliography{NGC6791-refs.bib} % your references Yourfile.bib

\end{document}